# Scaling Relations for the Cosmological "Constant" In Five-Dimensional Relativity

**Paul S. Wesson[1] and James M. Overduin[2]**

[1]Department of Physics and Astronomy, University of Waterloo, Waterloo, ON, N2L 3G1, Canada.
Email: psw.papers@yahoo.ca
[2]Department of Physics, Astronomy and Geosciences, Towson University, Towson, MD, 21252, U.S.A. Email: joverduin@towson.edu

When the cosmological "constant" is derived from modern five-dimensional relativity, exact solutions imply that for small systems it scales in proportion to the square of the mass. However, a duality transformation implies that for large systems it scales as the inverse square of the mass.

## 1. Introduction

The cosmological constant as it appears in Einstein's general relativity has several puzzling aspects, and it is a serious problem to understand why its value as inferred from cosmology is much smaller than its magnitude as implied by particle physics. However, it has been known for a long time that the cosmological "constant" appears more naturally when the world is taken to be five-dimensional [1], and recently there has been intense work on modern versions of 5D relativity where the extra dimension is not compactified [2-4]. The purpose of the present note is to draw together various results in the literature which indicate that there may be simple scaling relations between values of the cosmological "constant" $\Lambda$ and the mass $m$ of the system concerned. Tentatively, we identify $\Lambda \sim m^2$ for small systems and $\Lambda \sim 1/M^2$ for large, gravitationally-dominated systems. While these relations cannot be rigorously established with our present level of understanding, we believe that it is useful to point them out as guides for future research.

     The subjects which indicate possible relations are diverse, and include: the embedding of $\Lambda$-dominated solutions of 4D general relativity in the so-called 5D canonical metric [5-8]; the embeddings which lead to variable values of $\Lambda$ [9-13]; the equations of motion for canonical and related metrics [14-20]; conformal transformations which affect $\Lambda$ and possibly $m$ [21,22]; the vacuum and gauge fields associated with elementary particles [23,24]; and the wave-particle duality connected with certain $\Lambda$-dominated 5D metrics [25-27]. Most of our results are in Section 2. There we will re-examine the meaning of $\Lambda$, reinterpret two classes of known solutions, and present a new class with interesting properties. Section 3 is a conclusion.

     To streamline the working, we will often absorb the speed of light $c$, the gravitational constant $G$ and the quantum of action $h$, except in places where they are made explicit to aid understanding. As usual, upper-case Latin letters run $A,B = 0,123,4$ for time, space and the extra dimension. We label the last $x^4 = \ell$ to avoid confusion. Lower-case Greek letters run $\alpha,\beta = 0,123$. Other notation is standard.

## 2. The Cosmological "Constant" and Possible Scaling Relations

In this section, we will examine certain subjects which involve the cosmological "constant" $\Lambda$ of a spacetime and the mass $m$ of a test particle moving in it. That these parameters may be linked can be appreciated by noting that 5D relativity is broader than Einstein's 4D theory, being in general an account of gravity, electromagnetism and a scalar field, where the last is widely believed to be concerned with how particles acquire mass [2-4]. However, in 5D neither $\Lambda$ nor $m$ are in general constants. Rather, they depend on the field equations and solutions of them. It is common to take the field equations to be given in terms of the Ricci tensor by

$$R_{AB} = 0 \qquad (A,B = 0,123,4) . \qquad (1)$$

These apparently empty 5D equations actually contain Einstein's 4D equations with a finite energy-momentum tensor, a result guaranteed by Campbell's embedding theorem [5-7]. This means that the 4D theory is smoothly contained in the 5D one, and that the latter can be brought into agreement with observations at some level.

In Einstein's theory, the cosmological constant is usually introduced by adding a term $\Lambda g_{\alpha\beta}$ to the field equations:

$$G_{\alpha\beta} + \Lambda g_{\alpha\beta} = (8\pi G/c^4)T_{\alpha\beta} \qquad (\alpha,\beta = 0,123) . \qquad (2)$$

Here $g_{\alpha\beta}$ is the metric tensor, whose covariant derivative is zero, hence the acceptability of the noted term. We recognize that the $\Lambda$ term is a kind of gauge term. It is sometimes moved to the right-hand side of Einstein's equations, where it can be viewed as a vacuum fluid with density $\rho_v = \Lambda c^2/(8\pi G)$ and equation of state $p = -\rho_v c^2$. However, it should be recalled that the coupling constant between the left-hand (or geometrical) side of the Einstein equations and the right-hand (or matter) side is $8\pi G/c^4$. This therefore cancels the similar coefficient of the vacuum density, leading us back to the realization that $\Lambda$ is really a stand-alone parameter insofar as general relativity is concerned. (This is in line with the fact that its physical dimensions or units are $L^{-2}$, matching those of the rest of the field equations, which involve the second derivatives of dimensionless metric coefficients with respect to the coordinates.) An implication of this is that when $\Lambda$ is derived from a 5D as opposed to a 4D theory, it may be connected not with gravity but with the scalar field, a possibility we will return to later.

The quantum vacuum, as opposed to the classical one, is frequently attributed an energy density which is calculated in terms of many simple harmonic oscillators and expressed in terms of an effective value of $\Lambda$ [23]. This energy density is formally divergent, unless it is cut off by introducing a minimum wavelength, or equivalently a maximum wave-number $k$. With this understood, there results $\Lambda \sim \rho_v \sim hk^4/c$. If the cutoff in $k$ is chosen to be the inverse of the Planck length, this has size $\rho_v c^2 \sim 10^{112}$ erg cm$^{-3}$. For comparison, the cosmologically-determined value of $\Lambda$ ($\sim 10^{-56}$ cm$^{-2}$) corresponds to an energy density of order $10^{-8}$ erg cm$^{-3}$. The discrepancy, of order $10^{120}$, is the crux of the cosmological-constant problem.

An alternative interpretation of the result in the preceding paragraph is to imagine the quantum vacuum not spread through ordinary 3D space but concentrated in particles of mass $m$. It is reasonable to suppose that the stuff of each particle occupies a volume whose size is given by the Compton wavelength, $\lambda_c = h/mc$. Then the average density is approximately

$$\rho_v \sim m/\lambda_c^3 = hk^4/c \ . \tag{3}$$

This expression is formally identical to the one above. But the high-density vacuum is now confined to the particle, as expected if it is the product of a scalar field which couples to matter (see below). There is no conflict between (3) and the all-pervasive cosmological vacuum discussed above, so the cosmological-constant problem is avoided.

The best way to incorporate a scalar field into physics is to take its potential $\Phi$ to be the extra, diagonal element of an extended 5D metric tensor. Then following Kaluza the extra, non-diagonal elements can be identified with the potentials of electromagnetism, while the 4D block remains as a description of 4D Einsteinian gravity. Since we are here mainly interested in the scalar field, we can eliminate the electromagnetic potentials by a suitable use of the coordinate degrees of freedom of the metric, so the interval for the gravitational and scalar fields is

$$dS^2 = g_{\alpha\beta} dx^\alpha dx^\beta + \epsilon \Phi^2 d\ell^2 \ . \tag{4}$$

Here $g_{\alpha\beta}$ and $\Phi$ depend in general on both the coordinates of spacetime ($x^\gamma$) and the extra dimension ($\ell$). The symbol $\epsilon = \pm 1$ indicates whether the extra dimension is spacelike or timelike, both being allowed in modern 5D theory. (The extra dimension does not have the physical nature of an extra time, so for $\epsilon = +1$ there is no problem with closed timelike paths.) Many solutions are known of the field equations (1) for the metric (4) [2-4]. It transpires that the easiest way to approach the field equations is by splitting the 4D part of the metric into two functions, thus:

$$dS^2 = f(x^\gamma, \ell) \, \bar{g}_{\alpha\beta}(x^\gamma) \, dx^\alpha dx^\beta + \epsilon \Phi^2 d\ell^2 \ . \tag{5}$$

Here $f$ is a gauge function which determines the behavior in $x^4$, while $\bar{g}_{\alpha\beta}$ depends only on the spacetime coordinates $x^\gamma$. While the form (5) provides a mathematical advantage, it involves a physical quandary: does an observer experience the whole 4D space $ds^2 = f \bar{g}_{\alpha\beta} dx^\alpha dx^\beta$ or only the spacetime-dependent subspace $d\bar{s}^2 = \bar{g}_{\alpha\beta} dx^\alpha dx^\beta$? This question is akin to the argument for the so-called Jordan frame versus the Einstein frame in old 4D scalar-tensor theory, where a scalar function was applied to the 4D metric with no fifth dimension. It did not find a definitive answer then, and has not done so today. There is a difference in the physics between the two frames, but so long as the function $f$ is slowly varying, this will be minor. Cosmological observations may one day reveal the difference between the two frames, but for now we proceed with the view that they yield complementary physics.

An instructive case of the metric (5) has $f = (\ell/L)^2$ and $\Phi = 1$, where $\bar{g}_{\alpha\beta}$ is any solution of the Einstein equations without ordinary matter but with a vacuum fluid whose density is measured by $\Lambda$. This is known as the (pure) canonical metric. There is a large literature on this case (see Ref. [8] for a review). It includes the Schwarzschild-de Sitter metric for the Sun and solar system, and the de Sitter metric for the universe in its inflationary stage. It turns out that the equations of motion for a test particle in the 5D metric (5) are the same as those in the 4D theory, a result which enforces agreement with the classical tests of relativity [28,29]. The dynamics may be obtained either by using the 5D geodesic equation, or by putting $dS^2 = 0$ in (5). The latter is based on the fact that null paths in 5D with $dS^2 = 0$ reproduce the timelike paths of massive particles in 4D with $ds^2 > 0$, as well as the paths of photons with $ds^2 = 0$. The definition of dynamics and causality by $dS^2 = 0$ matches the null nature of the field equations (1). It turns out that the nature of the motion in the extra dimension $\ell = \ell(s)$ depends on the choice of $\epsilon$ in the metric (5), as does the sign of $\Lambda$. Thus introducing a constant $\ell_*$ we find

$$\ell = \ell_* \exp(\pm s/L), \qquad \Lambda = +3/L^2, \qquad \epsilon = -1 \qquad (6.1)$$

$$\ell = \ell_* \exp(\pm is/L), \qquad \Lambda = -3/L^2, \qquad \epsilon = +1. \qquad (6.2)$$

The second of these is of particular interest, because it is the same as the expression for the wave function $\psi = \psi_* \exp(\pm imcs/h)$ in old wave mechanics. In fact, it may be shown that the 5D geodesic equation for the (pure) canonical metric reproduces the Klein-Gordon equation with $\ell$ in place of $\psi$ and $1/L$ in place of $m$ [25-27]. We will meet the Klein-Gordon equation again below. Here we note that the (pure) canonical metric suggests the possibility that

$$|\Lambda| = 3/L^2 = 3(mc/h)^2. \qquad (7)$$

Here $m$ has been written in terms of the Compton wavelength. This identification presupposes that the observer experiences the 4D spacetime $\bar{g}_{\alpha\beta}$ in (5) rather than the composite spacetime defined by $f\bar{g}_{\alpha\beta}$. This is a subtle issue, as noted above, and we will return to it below.

The next most simple case of (5) is when a shift $\ell \to (\ell - \ell_0)$ is applied to the extra coordinate in the canonical metric. This may appear to be close to trivial, but is not because of the way in which the 4D Ricci scalar transforms, and with it $\Lambda$ [9,10,21,22]. The equations of motion and the mass of a test particle for the shifted canonical metric were worked out by Ponce de Leon [16-20]. He used the principle of least action and the eikonal equation for massive and massless particles, as opposed to the geodesic equation used by Mashhoon et al. [14,15]. As before, it turns out that $\Lambda > 0$ for a spacelike extra dimension ($\epsilon = -1$) and $\Lambda < 0$ for a timelike one ($\epsilon = +1$). The metric and the expressions for $\Lambda$ and $m$ are

$$dS^2 = [(\ell - \ell_0)/L]^2 \bar{g}_{\alpha\beta}(x^\gamma) dx^\alpha dx^\beta + \epsilon d\ell^2 \qquad (8.1)$$

$$|\Lambda| = 3/L^2 [\ell/(\ell - \ell_0)]^2 = 3(mc/h)^2. \qquad (8.2)$$

The second line here requires lengthy calculations for $\Lambda$ and $m$ [9,10,16-20], so the fact that we again find $|\Lambda| \sim m^2$ is significant.

The third case we present is more complicated than the canonical metrics studied in the two preceding paragraphs. In (5) we put $f = \exp(\ell\Phi/L)$ where $\Phi = \Phi(x^\gamma)$. This may be shown to satisfy the field equations (1), which break down into sets: ten relations which determine the energy-momentum tensor $T_{\alpha\beta}$ necessary to balance Einstein's equations; four conservation-type relations which fix a 4-tensor $P_{\alpha\beta}$ that has an associated scalar $P$; and one wave equation for the scalar field $\Phi$. The working is tedious (see Refs. [2-4]; indices are raised and lowered using $g_{\alpha\beta} = f\bar{g}_{\alpha\beta}$ of Eq. (5)). The metric and final results of the field equations read as follows:

$$dS^2 = \exp(\ell\Phi/L) \bar{g}_{\alpha\beta}(x^\gamma) dx^\alpha dx^\beta + \epsilon \Phi^2(x^\gamma) d\ell^2 \qquad (9.1)$$

$$8\pi T_{\alpha\beta} = \Phi_{,\alpha;\beta}/\Phi - \epsilon g_{\alpha\beta}/(2L^2) \tag{9.2}$$

$$P_\alpha{}^\beta = -3\delta_\alpha{}^\beta/(2L) \tag{9.3}$$

$$\Box\Phi + \epsilon\Phi/L^2 = 0. \tag{9.4}$$

Here a comma denotes the partial derivative, a semicolon denotes the (4D) covariant derivative, and $\Box\Phi \equiv g^{\alpha\beta}\Phi_{,\alpha;\beta}$ where $g^{\alpha\beta} = \exp(-\ell\Phi/L)\,\bar{g}^{\alpha\beta}(x^\gamma)$.

There are scalar quantities associated with the above which are of physical interest. For example, $T$ can be obtained by contracting (9.2) and using (9.4) to simplify it; $P$ as given by the contraction of (9.3) is a conserved quantity; and the (4D) Ricci or curvature scalar $R$ can be expressed in its general form and in the special form it takes for the metric (9.1). Thus:

$$8\pi T = \Box\Phi/\Phi - 2\epsilon/L^2 = -3\epsilon/L^2 \tag{10.1}$$

$$P = -6/L \tag{10.2}$$

$$R = (\epsilon/4\Phi^2)\,[g^{\alpha\beta}{}_{,4}\,g_{\alpha\beta,4} + (g^{\alpha\beta}\,g_{\alpha\beta,4})^2] = 3\epsilon/L^2. \tag{10.3}$$

These relations and (9) can be given physical interpretations along the lines of what has been done for other solutions in the literature [2-4]. The energy-momentum tensor (9.2) shows that the source consists of the scalar field plus a term which, because of its proportionality to $g_{\alpha\beta}$, would usually be attributed to a vacuum fluid with cosmological constant $\Lambda = -\epsilon/(2L^2)$. The conserved tensor of (9.3) obeys $P^\beta{}_{\alpha;\beta} = 0$ by the field equations, and its scalar $P$ has in other work been linked to the rest mass of a test particle, which here is $m = 1/L$ [25-27]. This is confirmed by the wave equation (9.4), which deserves some discussion.

Relation (9.4), depending on the choice for $\epsilon = \pm 1$, is known either as the Helmholtz equation or as the Klein-Gordon equation. Many solutions to it are known with applications to problems in atomic physics (like diffusion) and elementary particle physics (like wave mechanics). There are different modes of behavior, depending on whether $\epsilon = -1$ or $\epsilon = +1$, which correspond to the monotonic and oscillatory modes (6.1) and (6.2) of the canonical metric discussed before. For the present metric (9.1), the scalar field $\Phi$ may be real or complex, and in the latter case for $\epsilon = +1$ the wave equation (9.4) is identical to the Klein-Gordon equation, with $L = h/mc$ being the Compton wavelength of the test particle. This is similar to a previous interpretation based on the shifted-canonical metric [25-27]. (In (9.4), the oscillation is in $\Phi$, whereas in the corresponding equation of Refs. [25-27] it is in $\ell$, because in the canonical metric it is presumed that $\Phi = 1$, so the physical behavior is moved from the one parameter to the other. The problem can be made explicitly complex in (9.1) by writing $\exp(i\ell\Phi/L)$, if so desired.)

It may seem strange that a classical field theory yields an equation typical of (old) quantum theory, but it should be recalled that the wave equation (9.4) comes from the field equation $R_{44} = 0$, which does not exist in standard general relativity. In fact, the present interpretation of the metric (9.1) is fully consistent with the approach to noncompactified 5D relativity known as Space-Time-Matter theory, where matter on the macroscopic and microscopic scales is taken to be the result of higher-dimensional geometry [2-4]. By contrast, while the metric (9.1) may resemble the warp metric of the alternative approach to 5D relativity known as Membrane theory, in that approach the "Φ" in the exponent of the 4D part of the metric is absent, which means that the metric does not satisfy the field equations in the simple form (1). Our view is that (9) shows the wave-mechanical properties of matter. The scalars (10) associated with the solution bear this out. With conventional units restored, the conserved quantity $P$ is inversely proportional to the Compton wavelength $L = \lambda_c = h/mc$ of a test particle moving in the spacetime. Viewed as a wave which couples to matter, we expect that the Compton wavelength should be consistent with the radius of curvature of the spacetime, and this is confirmed by the relation for $R$. Lastly, we note that the aforementioned relation $\Lambda = -\epsilon/(2L^2)$ shows once again that $\Lambda \sim m^2$.

This relation is common to the three classes of solution examined above, which come from different choices of the gauge function $f(x^\gamma, \ell)$ in (5). They involve $f = (\ell/L)^2$ which gives (6), $f = [(\ell - \ell_0)/L]^2$ which gives (8), and $f = \exp(\ell\Phi/L)$ which gives (9). By comparison with known physics, we infer that the constant length $L$ is inversely proportional to the particle mass $m$, which we can write in terms of the Compton wavelength as $L \sim \lambda_c = h/mc$. The exponential gauge, in particular, leads from the field equation $R_{44} = 0$ to the Klein-Gordon equation, which is the basic relation in wave mechanics (its low-energy limit is the Schrödinger equation which underlies the physics of the hydrogen atom). The implication is that the scalar field of 5D relativity is connected to the mass of a particle, and with the phenomenon of wave-particle duality ([25-27]; the Klein-Gordon equation can have real or complex forms). These comments are in accordance with the longstanding view that theories of Kaluza-Klein type provide a way of unifying the interactions of particles with gravity. What, however, of the latter interaction? It is natural to wonder if there is not a complementary relation to what we have found above, but for macroscopic gravity-dominated systems.

This subject will require detailed analysis, but some comments of a preliminary type may be made. It is useful, in this context, to reconsider the traditional distinction between inertial mass ($m_i$) and gravitational mass ($m_g$). The Kaluza-Klein equation involves the former, so our previous considerations have concerned $L = h/m_i c$ and $\Lambda \sim m_i^2$ as the scaling relation for the cosmological constant. It is clear that this scaling rule cannot persist to arbitrarily large masses without leading to excessive curvature of empty spacetime ($R = 4\Lambda$). We expect, therefore, that it might pass over to some other scaling relation $\Lambda = \Lambda(m_g)$ for large gravitational masses.

Such a relation is actually implicit in certain work on the canonical metric [2-4,8-22]. We recall that the 4D part of the 5D canonical metric involves the combination $(\ell/L)ds$. This can be compared to the element of action for classical mechanics, $mds$. Two obvious identifications are possible: $L \sim 1/m$ and $\ell \sim m$. We have already explored the former, so attention focuses on the latter. In fact the possibility $x^4 \sim \ell \sim m_g$ has been considered, mainly in relation to cosmology, and cannot be ruled out [2-4,11-13]. As regards $\Lambda$, we note that its behavior depends on the coordinate frame experienced by an observer (see above). To illustrate this, consider a vacuum spacetime with the (pure) canonical metric, where the 4D part of the interval is $g_{\alpha\beta} dx^\alpha dx^\beta = (\ell/L)^2 \bar{g}_{\alpha\beta}(x^\gamma) dx^\alpha dx^\beta$. The effective value of $\Lambda$ can be obtained from either the Ricci scalar or the Einstein tensor, and depends on whether the observer experiences only $\bar{g}_{\alpha\beta}(x^\gamma)$ or the full $g_{\alpha\beta}(x^\gamma, \ell)$. The results are respectively $3/L^2$ and $3/\ell^2$, and both appear in the literature. Let us take the second alternative, and combine it with the physical identification $\ell \sim m_g$ noted above. The obvious parameter with which to geometrize the gravitational mass is $Gm_g/c^2$, the Schwarzschild radius. Then we find, in total, $\Lambda = 3(Gm_g/c^2)^{-2}$. That is, for large gravitationally-dominated systems we expect $\Lambda$ to scale as the inverse square of the mass.

The argument of the preceding paragraph is tentative, but can be checked by combining it with the more detailed work concerning the inertial mass which went before. For simplicity, we take numerical factors to be those of the canonical case, and consider a proton (inertial mass $m_p$) and the observable part of the universe (gravitational mass $M_u$). Then the scaling relations for the cosmological "constant" read $\Lambda_p = 3(m_p c/h)^2$ and $\Lambda_u = 3(c^2/GM_u)^2$. These can be combined to give the number of baryons in the observable universe as

$$N = M_u/m_p = (3c^3/Gh) / (\Lambda_p \Lambda_u)^{1/2} . \tag{11}$$

In this we substitute the quantum field theoretical value of $\Lambda_p = 2 \times 10^{26}$ cm$^{-2}$ and the cosmological value of $\Lambda_u = 3 \times 10^{-56}$ cm$^{-2}$ [obtained from $\Lambda_u = 8\pi G\rho_v/c^2$ where $\rho_v = \Omega_\Lambda \rho_{crit}$ and $\rho_{crit} = 3H_0^2/8\pi G$, together with current observational data giving $H_0 = 74 \pm 2$ km s$^{-1}$ Mpc$^{-1}$ and $\Omega_\Lambda = 0.73 \pm 0.04$]. The result is $N = 10^{80}$, in agreement with conventional estimates.

The two scaling relations considered in this section should be regarded as complementary. The first is better based in theory than the second, since it can be examined in three gauges rather than one. However, there is in principle no conflict between them, and in practice we expect the first to grade into the second. The $\Lambda \sim m^2$ rule should be dominant on the particle scale (~$10^{-13}$ cm), and the $\Lambda \sim 1/m^2$ rule should be dominant on the cosmological scale (~$10^{28}$ cm). Theoretically, they should be comparable on scales of order 100 km, which in practice is roughly where quantum interactions and solid-state forces are superseded by the effects of gravity.

## 3. Conclusion

We have seen in the preceding section that the cosmological constant is open to re-interpretation, particularly as a measure of the energy density of the vacuum fields of particles. It is somewhat better understood in cosmology, where its theoretical status is relatively clear in Einstein's equations, and where observations establish its approximate value. Unfortunately, there is a very large mismatch between the microscopic and the macroscopic domains. This can in principle be alleviated by using a five-dimensional theory, of the kind indicated by unification, where in general $\Lambda$ is not a universal constant but a variable. This is shown most clearly by the 5D canonical gauge, where $\Lambda$ scales according to the size of the potential well ($L$) or the value of the extra coordinate ($\ell$). Since the mass ($m$) of a test particle also depends on these parameters, we are tentatively led to suggest scaling relations of the form $\Lambda = \Lambda(m)$. For the canonical gauge in its pure and shifted forms, the scaling relation is for small $m$ and has the form $\Lambda \sim m^2$. This is also the form derived from the exponential gauge, which has the advantage of showing that the extra field equation resembles the Klein-Gordon equation of wave mechanics, implying that the scalar field is connected with particle mass.

There is, however, an alternative interpretation of the canonical gauge and others like it. The 4D part of this involves a term $(\ell/L)ds$, and to match the classical action $mds$ it is possible to use the gravitational mass with $\ell = Gm/c^2$ rather than the inertial mass with $L = h/mc$. The implication is that when gravity is dominant, for large $m$, there is a scaling relation of the form $\Lambda \sim 1/m^2$. This macroscopic relation should be viewed as complementary to the microscopic one, the changeover occurring at a length scale of order 100 km. When the two relations are combined, it is possible to obtain an expression for the number of baryons in the observable universe. This result (11) agrees with conventional estimates, which may be seen as provisional support for the idea that the cosmological "constant" varies with scale.

## Acknowledgment

Thanks for comments go to members of the Space-Time-Matter group (5Dstm.org).